\documentclass[12pt]{iopart}

\usepackage{iopams}
\usepackage[super,compress]{cite}
\usepackage{graphicx}
\usepackage{xcolor}

\expandafter\let\csname equation*\endcsname\relax

\expandafter\let\csname endequation*\endcsname\relax

\usepackage{amsmath}

\DeclareMathOperator\arctanh{arctanh}

\begin{document}
\title{Conformal Dilaton Gravity and Warped  Spacetimes in 5D}
\author{Reinoud Jan Slagter}

\address{Asfyon, Astronomisch Fysisch Onderzoek Nederland  \\and former\\
University of Amsterdam, The Netherland}

\ead{info@asfyon.com}
\vspace{10pt}
\begin{indented}
\item[]Dec 2020
\end{indented}
\begin{abstract}
An exact time-dependent solution of a black hole is found  in conformally invariant gravity on a warped  Randall-Sundrum spacetime, by writing the metric $g_{\mu\nu}=\omega^{\frac{4}{n-2}}\tilde g_{\mu\nu}$. Here $\tilde g_{\mu\nu}$ represents the "un-physical" spacetime and $\omega$ the dilaton field, which will be  treated on equal footing as any renormalizable scalar field.
In the case of a five-dimensional warped spacetime, we write $ ^{(4)}{\tilde g_{\mu\nu}}=\bar\omega^2{ ^{(4)}\bar g_{\mu\nu}}$
Both $\omega$ and $\bar\omega$ can be used to describe the different notion the in-going and outside observers have of the Hawking radiation by using different conformal gauge freedom. The disagreement about the interior of the black hole is explained by the antipodal map of points on the horizon. 
The free parameters of the solution can be chosen in such a way that $\bar g_{\mu\nu}$ is singular-free and topologically regular, even for $\omega\rightarrow 0$.
It is remarkable that the 5D and 4D effective field equations for the metric components and dilaton fields can be written in general dimension $ n= 4,5$.
It is conjectured that, in context of quantization procedures in the vicinity of the horizon, unitarity problems only occur in the bulk at large extra-dimension scale.
The subtraction point in an effective theory will be in the UV only in the bulk, because
the use of a large extra dimension results in a fundamental Planck scale comparable with the electroweak scale. 
\end{abstract}
\vspace{2pc}
\section{Introduction}
Our universe could be a 4-surface ("brane") embedded in a 5-dimensional spacetime, where the extra dimension represents the "bulk". We will focus on the "warped variant of these so-called brane-worlds gravity  models, originally introduced by Randall and Sundrum (RS)\cite{randal1,randal2} and extended by Shiromizu, et al.\cite{shirom1}. See also the overview work of Maartens\cite{roy1}.
At the same time, Arkani-Hamed, et al. (ADD)\cite{ark1,ark2} considered also  higher dimensional models which put forward the idea that a large volume for the compact extra dimensions will lower the Planck scale. Their models are not restricted to 5 dimensions, as is the case in the RS model with one "preferred" extra dimension. 
The advantage of these models is that they could solve the hierarchy problem, i.e., why there is such a huge gap between the electroweak and Planck scale. It has also promising results   for achieving late-time acceleration of the universe without dark energy or matter (DGP models)\cite{dvali1}.However, one should be careful in changing general relativity theory (GRT). If one adapt that Nature will choose the most simple Einstein-Hilbert (EH) action, then, for example, the cosmological constant should not be included in the EH action.  One avoids in this way the huge discrepancy between the contribution from zero-point fluctuations in quantum field theory and the predicted value in GRT (some 120 orders of magnitude). 

Another unsolved issue, is the complementarity near a black hole. 
In the vicinity of the horizon, curvature will be huge, so quantum effects will become important. Up till now, no convincing theory of quantum gravity  is available.
Many attempts were made in order to make a renormalizable and unitary quantum gravity model. I refer to the standard works on these subjects, for example Parker, et al\cite{parker} and the fundamental work of 't Hooft, et al.\cite{veltman}.

A well know method in constructing a renormalizable model\cite{stelle}, is to add fourth order derivative terms in the curvature tensor (Euler-term). However, one looses unitarity.
Also the "old" effective field theory (EFT) has its problems. One ignores  what is going on at high energy.
In order to solve the anomalies one encounters in calculating the effective action, one can apply the so-called conformal dilaton gravity (CDG) model\cite{thooft1,thooft2,thooft3,thooft4,duff,codello,alvarez}.
CDG is a promising route to tackle the problems arising in quantum gravity model, such as the loss of unitarity close to the horizon. 
One assumes local conformal symmetry, which is spontaneously broken (for example by a quartic self-coupling of the Higgs field). Changing the symmetry of the action was also successful in the past, i.e., in the SM of particle physics. 
A numerical investigation of the coupling was recently done\cite{slagter3}.
The key feature  in CDG, is the splitting of the metric tensor $g_{\mu\nu}=\omega^{\frac{4}{n-2}}(x)\tilde g_{\mu\nu}$, with $\omega$ the dilaton field. Applying perturbation techniques (and renormalization/dimensional regularization), in order to find the effective action and its divergencies, one first integrate over $\omega$ (shifted to the complex contour), considered as a conventional renormalizable scalar field and afterwards over $\tilde g_{\mu\nu}$ and matter fields.
The dilaton field is locally unobservable. It is fixed when we choose the global spacetime and coordinate system. If one applies this principle to a black hole spacetime, then the energy-momentum tensor of $\omega$ determines the Hawking radiation. We shall see that it can be calculated in our model.
When $\tilde g_{\mu\nu}$ is flat, then the handling of the anomalies  simplifies considerably\cite{thooft6}. When $\tilde g_{\mu\nu}$ is non-flat, the problems are more deep-seated.

In this context, the modification of GRT by an additional spacetime dimension could be an alternative compromise, because Einstein gravity on the brane will be modified by the very embedding  itself and opens up a possible new way to address the dark energy problem\cite{mann1,mann2}.
These models can be  applied to the standard Friedmann-Lemaitre-Robertson-Walker (FLRW) spacetime and the modification on the Friedmann equations can be investigated\cite{slagter1}.
Recently, Maldacena, et al.\cite{mald}, applies the RS model to two black hole spacetimes and could construct a traversable macroscopic wormhole solution by adding only a 5D U(1) gauge field (see also Maldacena\cite{mald2}).
However, an empty bulk would be preferable. In stead, one can investigate the contribution of the projected 5D Weyl tensor on the 4D brane. It carries information of the gravitational field outside the brane. 
If one writes the 5D Einstein equation in CDG setting, it could be possible that  an effective theory can be constructed  without an UV cutoff, because the fundamental scale $M_{5}$ can be much less than the effective scale $M_{Pl}$ due to the warp factor. The physical scale is therefore not determined by $M_{Pl}$

In section 2 we review the 5D RS model. In section 3 we outline the principles of CDG. In section 4 we apply the model to a black hole spacetime. In section 5 we present a physical interpretation.
All the calculations were done by computer-aided software.\footnote{the interested reader can obtain from the author the Maple file with all the calculations used in this manuscript}
\section{Brane-world models}

\subsection{The Randall-Sundrum model}
In the brane-world scenario under consideration (the RS1 model), the 4-dimensional world is described by a domain wall $({^{(4)}{\cal M}},{^{(4)}g_{\mu\nu}})$ in a 5-dimensional spacetime $({^{(5)}{\cal M}},{^{(5)}g_{\mu\nu}})$.  All the standard model fields are confined to the brane. On the other hand, gravity can propagate in the full spacetime. 
The Planck scale is now an effective coupling constant, describing gravity on scales much larger than the extra dimension ($d$) and related to the fundamental scale via the volume of the extra dimension $M_{pl}^2\sim M_{4+d}^{2+d} l^d$, with L the the length scale of the extra dimension.
If $L\sim \frac{1}{M_{pl}}$, then $M_{4+d}\sim M_{pl}$. If the extra dimensional volume is far above the Planck scale, then the true fundamental scale $M_{4+d}$  can be much less than the effective scale $\sim 10^{19} GeV$. From a different point of view, one could say that the dynamics of the higher dimensional gravitational fields influence the brane dynamics.
We used the notations of Shiromizu, et al.\cite{shirom1} and Maartens\cite{roy1}.
The 5-dimensional Einstein equations are
\begin{equation}
{^{(5)}}{G_{\mu\nu}}=-\Lambda_5{^{(5)}g_{\mu\nu}}+\kappa_5^2{^{(5)}T_{\mu\nu}},\label{2-1}
\end{equation}
with $\kappa_5^2=8\pi G_5=\frac{8\pi}{M_5^3}$. The negative bulk tension $ \Lambda_5$ is offset by the positive brane tension $\lambda$ (vacuum energy).
One writes the 5D metric
\begin{equation}
{^{(5)}g_{\mu\nu}}={^{(4)}g_{\mu\nu}}+n_\mu n_\nu\qquad {^{(5)}ds^2}={^{(4)}g_{\mu\nu}}(x^\alpha,y)dx^\mu dx^\nu+\Gamma(y)dy^2,\label{2-2}
\end{equation}
with $n^\mu$ the unit normal to the brane and $y$ the extra dimension. 
The extrinsic curvature of $y=const$ surfaces is given by
\begin{equation}
K_{\mu\nu}={^{(4)}g_{\mu}^{\alpha}}{^{(5)}}\nabla_\alpha n_\nu.\label{2-3}
\end{equation}
The Gauss-Codazzi equation determines the change of $K_{\mu\nu}$ along $y=const.$
\begin{equation}
{^{(4)}\nabla}_\mu K^\mu_\nu-{^{(4)}\nabla}_\nu K={^{(5)}R_{\mu\sigma}}{^{(4)}{g_\nu^\mu}}n^\sigma\label{2-4}
\end{equation}
The relation between the 4D and 5D Ricci tensors is given by the Gauss equation
\begin{equation}
{^{(4)}R_{\mu\nu}}={^{(5)}{R_{\rho\sigma}}}{^{(4)}}g_\mu^\rho{^{(4)}}g_\nu^\sigma-{^{(5)}{R^\alpha_{\beta\gamma\delta}}}n_\alpha{^{(4)}}g_\mu^\beta n^\gamma{^{(4)}}g_\nu^\delta
+KK_{\mu\nu}-K_\mu^\alpha K_{\nu\alpha}.\label{2-5}
\end{equation}
The second term on the  right-hand side is Eq.(\ref{2-1}) is zero, if there are no matter fields in the bulk $({^{(5)}T_{\mu\nu}}=0$). 
Using the Israel's junction conditions and by imposing $Z_2$-symmetry, one obtains the effective induced Einstein equations on the brane\cite{shirom1,roy1}
\begin{equation}
{^{(4)}G_{\mu\nu}}=-\Lambda_{eff}{^{(4)}g_{\mu\nu}}+\kappa_4^2{^{(4)}T_{\mu\nu}}+\kappa_5^4{\cal S}_{\mu\nu}+\frac{2}{3}\kappa_5^2{\cal F}_{\mu\nu}-{\cal E}_{\mu\nu},\label{2-6}
\end{equation}
with
${^{(4)}T_{\mu\nu}}$ the energy-momentum tensor on the brane, ${\cal S}_{\mu\nu}$ the quadratic contribution of the energy-momentum tensor ${^{(4)}T_{\mu\nu}}$ arising from the extrinsic curvature terms in the projected Einstein tensor,
\begin{equation}
{\cal S}_{\mu\nu}=\frac{1}{12}{^{(4)}}T{^{(4)}}T_{\mu\nu}-\frac{1}{4}{^{(4)}}T_{\mu\alpha}{^{(4)}}T{^\alpha_\nu}+\frac{1}{24}{^{(4)}g_{\mu\nu}}\Bigl(3{^{(4)}}T_{\alpha\beta}{^{(4)}}T^{\alpha\beta}-{^{(4)}}T^2\Bigr)\label{2-7}
\end{equation}
and ${\cal F}_{\mu\nu}$ is the correction term from the bulk energy-momentum tensor,
\begin{equation}
{\cal F}_{\mu\nu}={^{(5)}T_{\alpha\beta}}{^{(4)}g_\mu^\alpha}{^{(4)}g_\nu^\beta}+{^{(4)}g_{\mu\nu}}\Bigl({^{(5)}T_{\alpha\beta}}n^\alpha n^\beta-\frac{1}{4}{^{(5)}T}\Bigr).\label{2-8}
\end{equation}
In the RS model, the effective cosmological constant is $\Lambda_{eff}=\frac{1}{2}(\Lambda_5+\frac{\kappa_5^4}{6}\lambda^2)$, with $\lambda$ the brane tension, $\sim \frac{M_{pl}^2}{l^2}$
and $l$ the effective size of the extra dimension. Because there is no deviation of Newton's law down to  ${\cal O}(10^{-1})$, we have $\lambda>(1 TeV)^4, M_5>10^5 TeV$. The 4 dimensional cosmological constant is zero, when the RS balance between $\Lambda_5$ and $\lambda$ is broken.
The fundamental gravity scale is lowered, possibly even down to the electroweak scale $\sim TeV$.
For an empty bulk, ${\cal F}_{\mu\nu}=0$. 
The last term in Eq.(\ref{2-6}) represents the projected Weyl term and is a part of the 5D Weyl tensor that carries information of the gravitational field outside the brane.
\begin{equation}
{\cal E}_{\mu\nu}={^{(5)}}C^\alpha_{\beta\rho\sigma}n_\alpha n^\rho {^{(4)}}g_{\mu}^{\beta}{^{(4)}}g_{\nu}^{\sigma}.\label{2-9}
\end{equation}
Note that ${^{(4)}T_{\mu\nu}}$ is the energy-momentum tensor of the particles and fields confined to the brane (plus the term $\lambda {^{(4)}}g_{\mu\nu}$).
The extrinsic curvature of the brane can then be expressed as
\begin{equation}
K_{\mu\nu}=-\frac{1}{2}\kappa_5^2\Bigl({^{(4)}T_{\mu\nu}}-\frac{1}{3} {^{(4)}}T\Bigr).\label{2-10}
\end{equation}
The 4D contracted Bianchi identity becomes (for empty bulk)
\begin{equation}
{^{(4)}}\nabla^\mu {\cal E}_{\mu\nu}=\kappa_4^2{^{(4)}\nabla^\mu T_{\mu\nu}}+\kappa_5^4\nabla^\mu{\cal S}_{\mu\nu}.\label{2-11}
\end{equation}
It relates ${^{(4)}}\nabla^\mu {\cal E}_{\mu\nu}$ to the several "matter terms" on the right-hand side of Eq.(\ref{2-11}) and expresses the fact that 4D spacetime variation in the matter-radiation can source KK-modes. 
We shall see in an application in section 4, that this identity will lead to a third-order partial differential equation, which forms with the effective Einstein equations, a closed system of pde's.
Further, Eq.(\ref{2-4}) becomes
\begin{equation}
{^{(4)}\nabla}_\mu K^\mu_\nu-{^{(4)}\nabla}_\nu K=\kappa_5^2{^{(4)}T_{\rho\sigma}}n^\sigma{^{(4)}}g_\nu^\rho.\label{2.12}
\end{equation}
One must realize that in the conformal dilaton application,  ${^{(4)}T_{\mu\nu}}$ contains also a contribution from the dilaton field $\omega$, which is a "scale"-term by the splitting $g_{\mu\nu}=\omega^2\tilde g_{\mu\nu}$ (section 3). The dilaton will play the role of a "scalar field". In general, the covariant splitting which is applied in the warped spacetime model, has the advantage that it shows a clear insight into the interplay between the 4D and 5D gravitational fields.
However, one needs to boundary conditions on the brane for a consistent dynamical evolution.
\subsection{Application to the FLRW spacetime}
In a former study we investigated the brane-world model of Randall and Sundrum on the warped 5 dimensional spacetime\cite{slagterpan}
\begin{equation}
ds^2=W(t,r,y)^2\Bigl[e^{2\gamma(t,r)-2\psi(t,r)}(-dt^2+dr^2)+e^{2\psi(t,r)}dz^2+\frac{r^2}{e^{2\psi(t,r)}}d\varphi^2\Bigr]+\Gamma(y)dy^2.\label{2-13}
\end{equation}
In first instance, we will take $\Gamma(y)=1$.
From the 5D Einstein equations
\begin{equation}
^{(5)}{G_{\mu\nu}}+\Lambda_5 ^{(5)}{g_{\mu\nu}}=0,\label{2-14}
\end{equation}
one obtains for the warpfactor
\begin{equation}
W(t,r,y)=W_1(t,r)W_2(y),\label{2-15}
\end{equation}
with
\begin{equation}
W_2(y)=e^{\sqrt{-\frac{1}{6}\Lambda_5}(y-y_0)}\label{2-16}
\end{equation}
and
\begin{equation}
W_1(t,r)^2=\pm\frac{1}{\tau r}\Bigl(d_1e^{\sqrt{2\tau}t}-d_2e^{-\sqrt{2\tau}t}\Bigr)\Bigl(d_3e^{\sqrt{2\tau}r}-d_4e^{-\sqrt{2\tau}r}\Bigr),\label{2-17}
\end{equation}
where $\tau$ and $d_i$ are constants. One recognizes the well-known RS warpfactor $W_2(y)$.
It turns out that $W_1$ cannot be isolated from the effective 4D brane equations.
\subsection{Application to a black hole spacetime}
Let us now apply the equation on the spacetime (a 4D BTZ model\cite{banadoz,carlib,compere,slagter4,slagter1})
\begin{equation}
ds^2=W(t,r,y)^2\Bigl[-N(t,r)^2dt^2+\frac{1}{N(t,r)^2}dr^2+dz^2+r^2(d\varphi+N_\varphi(t,r)dt)^2\Bigr]+dy^2.\label{2-18}
\end{equation}

Again, from the 5D Einstein equations, one obtains for $W(t,r,y)=W_1(t,r)W_2(y)$, with $W_2(y)$ the same expression, i. e., Eq.(\ref{2-16}). However, $W_1(t,r)$ cannot be isolated. One needs the 4D effective Einstein equations. It will become clear, that $W_1$ is related to the scale of spacetime of the geometry  near horizon region.
In order to make the connection with the CDG model, one would like to  write the metric as
\begin{equation}
ds^2=W(t,r,y)^2\Bigl[-N(t,r)^2dt^2+\frac{1}{N(t,r)^2}dr^2+dz^2+r^2(d\varphi+N_\varphi(t,r)dt)^2+dy^2\Bigr].\label{2-19}
\end{equation}
From the 5D Einstein equations, however, we then  obtain that $W_2(y)=const.$ 
\section{Conformal dilaton gravity}
\subsection{The basics}
Where Einstein's theory of general relativity treats the large-scale behavior of gravity, the Standard Model (SM) of particle physics is the theoretical framework where, on small-scale, the other three forces  of nature are explained. Combining these two theories leads to fundamental difficulties regarding the nature of spacetime. It is one of the greatest challenges  of theoretical physics. It is believed that the unification has taken place  close to the Planck scale, $1.6 .10^{-35} m$, so far out of reach of experimental verification. 
It is conjectured, however, that close to the horizon of a black hole, quantum gravity effects will be manifest. So observations on black holes could be a way to collect quantum-gravitational information.
It is well known that Einstein gravity, when quantized, is unitary, but nonrenormalizable\cite{veltman}.
In general, there is no obvious way to attain renormalizability without violating unitarity.
Higher order derivative theories turn out to be renormalizable at the one-loop quantum level, but lacks unitarity of the S-matrix. In the low energy domain ( well below the UV cutoff), one can construct an effective field theory that can be quantized and is renormalizable. However, just as in Standard Model (SM), one needs gauge-fixing terms as well as Faddeev-Popov ghost fields.
However, this route runs into problems when one tries to incorporate dark mass, dark energy and singularities. 
CDG could offer an alternative. Conformal gravity is part of the covariant approach to quantum gravity.
There are several methods to attack the consequences of conformal symmetyry at the quantum level. We follow the method initiated by 't Hooft\cite{thooft1,thooft2,thooft3}. See also Alvarez, et al.\cite{alvarez} and Codello, et al.\cite{codello}
The fundamental calculation of the divergences of the action involving the gravitational field, where already done long time ago\cite{veltman,stelle}. See also the overview work of Parker\cite{parker}.
Let us summarize the CDG  idea\cite{thooft1,thooft2}.
One can split the spacetime (in the 5D case)
\begin{equation}
^{(5)}{g_{\mu\nu}}=\omega^{4/3}{^{(5)}{\tilde g_{\mu\nu}}},\label{3-1}
\end{equation}
where $\omega$ is a dilaton metascalar, encoding all the scale dependencies. One can call the "un-physical" $\tilde g_{\mu\nu}$ a metatensor, i.e., it transforms as a tensor, but contains powers of the Jocobian of the coordinate transformation. One can easily prove that the action (in n dimensions and without matter terms)
\begin{eqnarray}
S=\int d^nx\sqrt{-\tilde g}\Bigl[\frac{1}{2}\xi \omega^2\tilde R+\frac{1}{2}\tilde g^{\mu\nu} \partial_\mu\omega\partial_\nu\omega+\Lambda\kappa^{\frac{4}{n-2}}\xi^{\frac{n}{n-2}}\omega^{\frac{2n}{n-2}} \Bigr]\label{3-2}
\end{eqnarray}
is conformal invariant under
\begin{equation}
g_{\mu\nu}\rightarrow \Omega^{\frac{4}{n-2}} g_{\mu\nu},\quad \omega \rightarrow \Omega^{-\frac{n-2}{2}}\omega.\label{3-3}
\end{equation}
The coupling constant $\Lambda$ is dimensionless in any dimension. It fulfills also the off-shell Ward identity.
The constant $\xi=\frac{n-2}{4(n-1)}$. One re-scaled $\omega \rightarrow \kappa\sqrt{\xi}\omega$, with $\frac{1}{\kappa^2}=M_{pl}^{n-2}=\frac{1}{16\pi G_N}$. So Newton's constant has disappeared completely.
Variation of the action leads to the field equations
\begin{eqnarray}
\xi\omega\tilde R-\tilde g^{\mu\nu}\tilde\nabla_\mu\tilde\nabla_\nu\omega-\frac{2n}{n-2}\Lambda\kappa^{\frac{4}{n-2}}\xi^{\frac{n}{n-2}}\omega^{\frac{n+2}{n-2}}=0\label{3-4}
\end{eqnarray}
and
\begin{eqnarray}
\omega^2\tilde G_{\mu\nu}=T_{\mu\nu}^{\omega}-\Lambda \tilde g_{\mu\nu}\kappa^{\frac{4}{n-2}}\xi^{\frac{2}{n-2}}\omega^{\frac{2n}{n-2}},\label{3-5}
\end{eqnarray}
with
\begin{eqnarray}
T_{\mu\nu}^\omega=\tilde\nabla_\mu\tilde\nabla_\nu\omega^2-\tilde g_{\mu\nu}\tilde\nabla^2\omega^2+\frac{1}{\xi}\Bigl(\frac{1}{2}\tilde g_{\alpha\beta}\tilde g_{\mu\nu}-\tilde g_{\mu\alpha}\tilde g_{\nu\beta}\Bigr)\partial^\alpha\omega\partial^\beta\omega .\label{3-6}
\end{eqnarray}
The covariant derivatives are taken with respect to $\tilde g_{\mu\nu}$.
\subsection{Notes on anomalies}
In general, one conjectures that conformal invariance is an exact symmetry, comparable with  the Higgs mechanism. The effective action will stay locally conformal invariant after the $\omega$ integration. The integration over $\tilde g_{\mu\nu}$ (and matter terms) will not be performed before imposing conformal constraints\footnote{We have still the freedom of $\Omega$ (Eq.(\ref{3-3})). So different observers experience different $\bar \omega$ and also different matter distribution (note that the Ricci tensor is not invariant under Eq.(\ref{3-3}): $R\rightarrow \frac{1}{\Omega^2}R-\frac{6}{\Omega^3}\nabla^\mu\partial\Omega$). }.  
The problem is how to handle the functional integration over $\tilde g_{\mu\nu}$ (the "curved background anomaly"). 
On a flat background, the conformal anomalies can be handled properly\cite{thooft3}.
The choice of $\omega$ is then unique. $\omega$ is a renormalizable field which could create the spacetime twofold: an infalling and outside observer use different ways to fix the conformal gauge in order to overcome unitarity problems.
For non-flat $\tilde g_{\mu\nu}$, standard QM does not apply, because there is no notion of energy in a conformal invariant theory.
Further, if one adds matter sources to the Lagrangian, for example a scalar gauge field $(\Phi,A_\mu)$, it will break the conformal invariance, i.e., it spoils the tracelessness of the energy momentum tensor. However, if one shifts $\omega$ to the complex contour ($\omega^2\rightarrow -\frac{\omega^2}{\kappa^2}$), then the  scalar field and dilaton field can be handled on equal footing (in order to obtain the same unitarity properties).
Using dimensional regularization and renormalization, it was found that the scalar field Lagrangian
\begin{equation}
{\cal L}=\sqrt{-g}\Bigl(-\frac{1}{2}g^{\mu\nu}\partial_\mu\phi\partial_\nu\phi-\frac{1}{12}R\phi^2\Bigr),\label{3-7}
\end{equation}
generates an effective action, whose divergent part is 
\begin{equation}
\Gamma^{div}\sim\frac{\sqrt{-g}}{4-n}(R_{\mu\nu}R^{\mu\nu}-\frac{1}{3}R^2).\label{3-8}
\end{equation}
One can apply this result directly to the Lagrangian Eq.(\ref{3-2}), where $R$ and $g$ are replaced by $\tilde R$ and $\tilde g$.
So the effective action becomes
\begin{equation}
S^{eff}=C\int d^nx\frac{\sqrt{-\tilde g}}{8\pi^2(4-n)}\Bigl(\tilde R^{\mu\nu}\tilde R_{\mu\nu}-\frac{1}{3}\tilde R^2\Bigl).\label{3-9}
\end{equation}
We will treat now $\tilde g_{\mu\nu}$ classical, so we will have no unitarity problem. 
Further, we have still the infinite coefficient in front of the effective action. It is conjectured\cite{thooft1}, that no counter term is needed in order to make the integral finite. 
The divergent coefficient in front of $S^{eff}$ will not be a problem when  $\tilde g_{\mu\nu}$ is treated  classical: $N$ and $N^{\varphi}$ can take large values and in the limit that $\frac{C}{4-n} \rightarrow\infty$, quantum fluctuations would vanish and only the classical parts would remain. Only $\omega$ (and matter fields) are described by renormalizable Lagrangians.
This means that $\tilde g_{\mu\nu}$ is flat beyond the Planck scale, but gets renormalized by $\omega$ and matter fields at much lower scales.

Quantum amplitudes are obtained by integrating the exponent of the entire action over all the components of $g_{\mu\nu}$, now written as $\omega^2\tilde g_{\mu\nu}$ (or, in the 5D case, $^{(5)}{g_{\mu\nu}}=\omega^{4/3}{^{(5)}{\tilde g_{\mu\nu}}}$). The integration over $\omega$ must be along a complex contour\cite{thooft1}.
Further, we consider the brane spacetime  $^{(5)}{\tilde g_{\mu\nu}}=^{(4)}{\tilde g_{\mu\nu}}+n_\mu n_\nu$
with now $^{(4)}{\tilde g_{\mu\nu}} =\bar\omega^2{^{(4)}{\bar g_{\mu\nu}}}$. Our $\bar g_{\mu\nu}$ is given by the up-lifted BTZ black hole spacetime. The evolution equation for $N$ will contain no intrinsic mass or length scale. (for $\Lambda =0$) and reflects the fact that we are dealing here with gravitational waves only. The scale is determined by $\omega$ and $\bar\omega$\footnote{We shall see that in the vacuum case, we are dealing with two different scales, $\omega$ and $\bar\omega$. Note, however, 
that $\omega^{4/3}=\bar\omega^2$ as expected}.
One says that $\omega$ is part of the gravitational sector. It fixes the scale at which quantum gravity effects becomes important.

It is obvious, that close to the horizon and ergo-sphere of the spacetime Eq.(\ref{3-7}), the curvature is high and quantum effects will become important. 
We already mentioned  that the Lagrangian Eq.(\ref{3-2}) with $\Lambda=0$ contains no intrinsic mass or length scale. This is comparable with the classical Lagrangian of the SM, if one omits the Higgs mass term. So it is challenging to apply a spontaneous symmetry breaking mechanism of conformal invariance by adding a mass term $-\frac{1}{2}\sqrt{-\bar g}m^2\bar\omega^2\bar\phi^2$ (through the quartic self-coupling of the Higgs field) and other contributions\cite{thooft3}, just as the Higgs mechanism in SM. Newton's constant then reappears by the re-scaling $\omega\rightarrow \kappa\sqrt{\xi}\omega$. 
\footnote{For the renormalization problems caused by the mass term, we refer to the literature. It is conjectured that it will not spoil conformal invariance of the effective action. The application to our model is currently under investigation.}
One conjectures that the effective action (with matter fields and on the black hole spacetime))  will be  locally spontaneously broken. 
\cite{thooft6}

Now the RS idea can be applied: the renormalized counter terms  in $S^{eff}$ will cause unitarity  problems only in the bulk at the large extra-dimension-scale ($l$), but are invisible on the brane.
This means that the  substraction point will be in the UV region as seen by an observer in the bulk.
Note that this is possible, because  we have in the induced 4D  Einstein equations the KK-contribution from the bulk, i.e., the electric part of the Weyl tensor ${\cal E}_{\mu\nu}$. From the brane observer, these KK corrections are non-local, since they incorporate 5D gravity wave modes.
Note that if the extra dimensional volume is significantly above the Planck scale, than the true fundamental scale $M_{4+d}$  can be much less than the effective scale $M_{pl}\sim 10^{19} GeV$. In the RS-1 model, $M_5^3=\frac{M_{pl}}{l}$, with $l$ the effective length scale  scale of the infinite large extra dimension $y$. The warp factor (Eq.(\ref{2-16})) then causes a finite contribution to the 5D volume
\begin{equation}
\int d^5 x\sqrt{-^{(5)}g} \sim \int d^4x\int dy W_2\sim l\int d^4 x.\label{3-10}
\end{equation} 
So the effective size of the extra dimension probed by the 5D graviton, is $l$. Note that the negative bulk tension is taken $\Lambda_5\sim -\frac{1}{l^2}\sim -\mu^2$ in the RS-1 model, with $\mu$ the energy scale. So for low energy scale  it prevents gravity leaking into the extra dimension. As by-product, the model solves the hierarchy problem.\\
In order to solve the conformal anomaly, caused by the interaction of the matter fields and gravity, one calculates the beta-functions (note that the trace is proportional to the beta functions), which has to be zero (these finite quantum theories are scale invariant) For a renormalizable theory, there must be as many beta functions as adjustable parameters, thus allowing to completely determine the theory.
Another problem is the dynamics of $\bar g_{\mu\nu}$. If the behavior is classical, then there is an interaction with the quantum fields, i.e., the dilaton field.
On a black hole spacetime,  one has to deal with Hawking radiation, even in the absent of matter terms, by outgoing gravitational waves. This is represented by the dynamical equation for the dilaton field, which decouples from $\bar g_{\mu\nu}$. It is clear that $\bar\omega$ will not decouple when a scalar field is incorporated\cite{slagter3}.

We shall see  in section 4 that $\bar\omega$ determines the evolution of  $\bar g_{\mu\nu}$ (i.e., $N$ and $N^\varphi$).
\section{The field equations of the warped CDG model }
\subsection{The 5D equations}
Let us apply the  equations of section 3 to the 5D spacetime ($^{(5)}{g_{\mu\nu}}=\omega^{4/3} {^{(5)}{\tilde g_{\mu\nu}}}$) 
\begin{equation}
ds^2=\omega(t,r,y)^{4/3}\Bigl[-N(t,r)^2dt^2+\frac{1}{N(t,r)^2}dr^2+dz^2+r^2(d\varphi+N_\varphi(t,r)dt)^2+dy^2\Bigr].\label{4-1}
\end{equation}
This metric is used for the 3D BTZ black hole solution (where the $dz^2$ is removed) and can be  up-lifted to 4D\cite{slagter1,slagter3}.
From the 5D Einstein equations Eq.(\ref{3-5}), one obtains of course again  $\omega(t,r,y)=\omega_2(y)\omega_1(t,r)$, with $\omega_2(y)= l$ = constant = length scale of the extra dimension.


The set of PDE's for $\omega$ and $N$ now becomes (we omit the subscript 1 in $\omega$)
\begin{equation}
\ddot\omega=-N^4{\omega}''+\frac{5}{3\omega}(N^4{\omega '}^2+\dot\omega^2)\label{4-2}
\end{equation}
\begin{equation}
\ddot N=\frac{3\dot N^2}{N}-N^4\Bigl(N''+\frac{3N'}{r}+\frac{N'^2}{N}\Bigr)
-\frac{2}{\omega}\Bigl[N^5\Bigl(\omega''-\frac{5\omega'^2}{3\omega}+\frac{\omega'}{r}\Bigr)
+N^4\omega'N'+\dot\omega\dot N\Bigr]\label{4-3}
\end{equation}
The equation for $N$ can also be presented without $\omega''$ on the right hand side. Then there is an extra constraint equations in $\omega''$:
\begin{equation}
\omega''=-\frac{\Lambda\omega^{7/3}\kappa^{4/3}\sqrt[3]{18}l}{16N^2}-\frac{\omega' N'}{N}
-\frac{\omega'}{2r}+\frac{4\dot\omega^2}{3\omega N^4}-\frac{\dot\omega\dot N}{N^5}.\label{4-4}
\end{equation}
A non-trivial  solution for $\omega$  becomes (compare with the warped spacetime solution of Eq.(\ref{2-17})
\begin{equation}
\omega=\Bigl[\frac{c_1}{(r+c_2)t+c_3r+c_4}\Bigr]^{3/2}\label{4-5}
\end{equation}
(or $r$ and $t$ interchanged).
For $c_2c_3=c_4$, we find the exact solution for $N$
\begin{equation}
N^2=\frac{1}{5r^2}\frac{10c_2^3r^2+20c_2^2r^3+15c_2r^4+4r^5+D_1}{D_2(c_3+t)^4+D_3},\label{4-6}
\end{equation}
with $c_i, D_i$ constants. Note that $N$ can be written as $N=N_1(r)N_2(t)$.
The dilaton equation Eq.(\ref{3-4}) is superfluous. One easily checks that the trace of Eq.(\ref{3-5}) is zero and that $\tilde\nabla^\mu( T_{\mu\nu}^{\omega}-\Lambda \tilde g_{\mu\nu}\kappa^{\frac{4}{n-2}}\xi^{\frac{2}{n-2}}\omega^{\frac{2n}{n-2}} )=0$. Also, the Ward identity is fulfilled.
The equations for $N^\varphi$ are
\begin{equation}
{\dot{N^\varphi}}'=-\frac{2\dot\omega {N^\varphi}'}{\omega}\label{4-7}
\end{equation}
\begin{equation}
{{N^\varphi}'}^2=-\frac{4N}{r^3}\Bigl(N'+\frac{N\omega'}{\omega}\Bigr)\label{4-8}
\end{equation}
\begin{equation}
{N^\varphi}''=-{N^\varphi}'\Bigl(\frac{2\omega'}{\omega}+\frac{3}{r}\Bigr)\label{4-9}
\end{equation}
\begin{figure}[h]
\centerline{
\includegraphics[width=8cm]{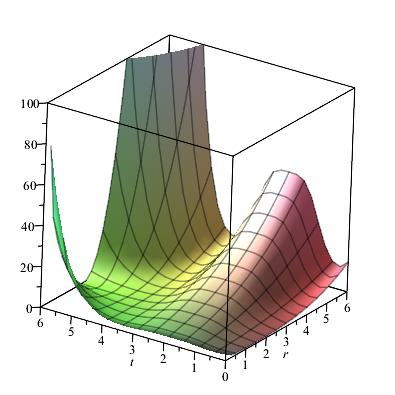}
\includegraphics[width=8cm]{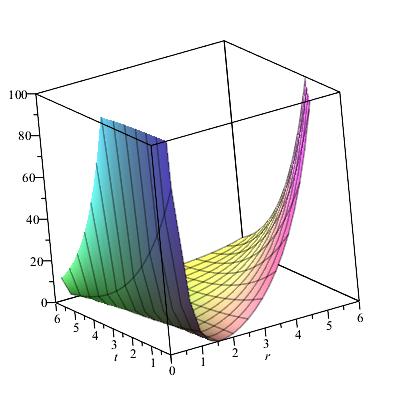}}
\caption{Example of an exact solution of $\tilde g_{tt}$ (left) and $\omega^{4/3}\tilde g_{tt}$}
\end{figure}
with solution
\begin{equation}
N^\varphi =F_1(t)+\int\frac{1}{r^3\omega^2}dr\label{4-10}
\end{equation}
with $F_1(t)$ an arbitrary function in t. 
A special solution of $N^\varphi$ is
\begin{equation}
N^\varphi =\frac{(t+c_3)^3}{2r^2}\Bigl(6c_2r^2 ln(r)-c_2^3-6c_2^2r+2r^3\Bigr)\label{4-11}
\end{equation}
Further, we have the constraint
\begin{equation}
N'=-\frac{N\omega'}{\omega}+\frac{e_1}{N\omega^4 r^3}\label{4-12}
\end{equation}
with $e_1$ a constant. This constraint can be used as boundary condition in the numerical solution.

In figure 1  we plotted a typical exact solutions  of $\omega^{4/3}g_{tt}$  and $g_{tt}$. The Killing  horizon is where   the ergo region begins.
Note however, that the general exact solution for $\omega$ has one more free parameter $c_4$.

\subsection{The effective 4D Einstein equations on the brane}
We now write
\begin{equation}
^{(5)}{\tilde g_{\mu\nu}}={^{(4)}}{\tilde g_{\mu\nu}} +n_\mu n_\nu\label{4-13}
\end{equation}
with $n_\mu=[0,0,0,0,\sqrt{l}]$ the unit normal to the brane. Ones again, we define
\begin{equation}
{^{(4)}}{\tilde g_{\mu\nu}}=\bar\omega^2 \bar g_{\mu\nu}\label{4-14}
\end{equation}

The effective Einstein equations on the brane are
\begin{equation}
{^{(4)}{\bar G_{\mu\nu}}}=\frac{1}{\bar\omega^2}\Bigl(-\Lambda_{eff}{^{(4)}{\bar g_{\mu\nu}}}-\Lambda\kappa^2 {^{(4)}{\bar g_{\mu\nu}}}\bar\omega^4+{^{(4)}}{T_{\mu\nu}}^{(\bar\omega)}\Bigr)-{\cal E}_{\mu\nu},\label{4-15}
\end{equation}
with ${\cal E}_{\mu\nu}$ the electric part of the Weyl tensor, given by
\begin{equation}
{\cal E}_{\mu\nu}={^{(5)}{C_{\alpha\beta\gamma\delta}}}n^\delta {^{(4)}{\bar g_\mu^\alpha}}{^{(4)}{\bar g_\nu^\beta}}\label{4-16}
\end{equation}
Note that ${\cal E}_{\mu\nu}$ is trace free.
For the "un-physical" metric ${^{(4)}}{\bar g_{\mu\nu}}$ we have now
\begin{equation}
ds^2=-N(t,r)^2dt^2+\frac{1}{N(t,r)^2}dr^2+dz^2+r^2(d\varphi+N_\varphi(t,r)dt)^2\Bigr].\label{4-17}
\end{equation}
Again, an equation for $\bar\omega$ can be isolated (we take from now on $\Lambda_{eff}=0$), i.e., 
\begin{equation}
\ddot{\bar\omega}=-N^4{\bar\omega}''+\frac{2}{\bar\omega}(N^4{\bar{\omega}}^{'2}+\dot{\bar \omega}^2)\label{4-18}
\end{equation}
\begin{eqnarray}
\ddot N=\frac{3\dot N^2}{N}-N^4\Bigl(N''+\frac{3N'}{r}+\frac{N'^2}{N}\Bigr)\cr
-\frac{3}{\bar\omega}\Bigl[N^5\Bigl(\bar\omega''-\frac{2\bar\omega'^2}{\bar\omega}+\frac{\bar\omega'}{r}\Bigr)
+N^4\bar\omega'N'+\dot{\bar\omega}\dot N\Bigr]
\label{4-19}
\end{eqnarray}
Now the non-trivial solution of $\bar\omega$ becomes (compare with Eq.(\ref{4-5}) )
\begin{equation}
\bar\omega=\Bigl[\frac{b_1}{(r+b_2)t+b_3r+b_4}\Bigr],\label{4-20}
\end{equation}
with $b_i$ constants. 
The solution for N is the same as Eq.(\ref{4-6}) (with $c_i$ replaced by $b_i$ and $D_i$ by $C_i$)
\begin{equation}
N^2=\frac{1}{5r^2}\frac{10b_2^3r^2+20b_2^2r^3+15b_2r^4+4r^5+C_1}{C_2(b_3+t)^4+C_3},\label{4-20b}
\end{equation}
The constraint for $\bar\omega''$ becomes now
\begin{equation}
\bar\omega''=-\frac{\Lambda\bar\omega^3\kappa^2l}{9N^2}-\frac{\omega' N'}{N}
-\frac{\omega'}{2r}+\frac{2\dot{\bar\omega}^2}{\bar\omega N^4}-\frac{\dot{\bar\omega}\dot N}{N^5}\label{4-21}
\end{equation}

The equations for $N^\varphi$ are
\begin{equation}
{\dot{N^\varphi}}'=-\frac{3\dot{\bar\omega} {N^\varphi}'}{\omega}\label{4-22}
\end{equation}
\begin{equation}
{{N^\varphi}'}^2=-\frac{2N}{r^3}\Bigl(2N'+3\frac{N\bar\omega'}{\bar\omega}\Bigr)\label{4-23}
\end{equation}
\begin{equation}
{N^\varphi}''=-{3N^\varphi}'\Bigl(\frac{\bar\omega'}{\bar\omega}+\frac{1}{r}\Bigr)\label{4-24}
\end{equation}
with solution
\begin{equation}
N^{\varphi}=F_2(t)+\int\frac{1}{r^3\bar\omega^3}dr\label{4-25}
\end{equation}
with $F_2(t)$ an arbitrary function in t. 
A typical solution is also given by Eq(\ref{4-11})!

Further, we have the constraint
\begin{equation}
N'=-\frac{3N\bar\omega'}{2\bar\omega}+\frac{e_2}{N\bar\omega^6 r^3}\label{4-26}
\end{equation}
with $e_2$ a constant.
\begin{figure}
\centerline{\includegraphics[width=9cm]{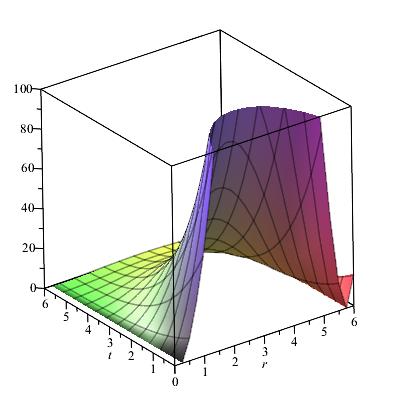}
\includegraphics[width=9cm]{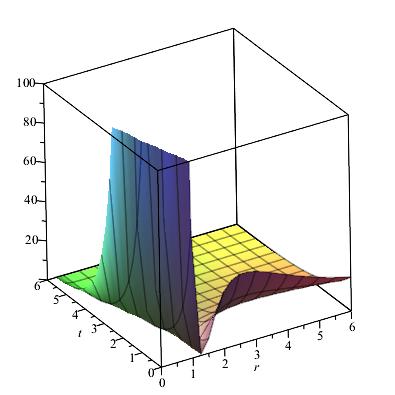}}
\caption{Left: typical plot of  $\bar g_{tt}$  for the 4D induced spacetime. Right: typical plot of $\omega^{4/3}\bar\omega^2\bar g_{tt}$ . We observe that the ergo sphere fades away as time increases}.  
\end{figure}
In figure 2 we plotted  $g_{tt}$ for  $\bar g_{\mu\nu}$ and $g_{\mu\nu}$.

Again, the dilaton equation is superfluous and the trace of Eq.(\ref{3-5})  again zero. 
The conservation equations now become
\begin{equation}
\bar\nabla^\mu{\cal E}_{\mu\nu}=\bar\nabla^\mu\Bigl[\frac{1}{\bar\omega^2}\Bigl(-\Lambda\kappa^2{^{(4)}{\bar g_{\mu\nu}}}\bar\omega^4+{^{(4)}}{T_{\mu\nu}}^{(\bar\omega)}\Bigr)\Bigr]\label{4-27}
\end{equation}
which yields differential equations for $\ddot{N'}$ and $\dot N$ as boundary conditions at the brane.
It can be described as the non-local conservation equation.
In the high energy case close to the horizon, one must include the ${\cal S}_{\mu\nu}$ term. 
So the divergence of ${\cal E}_{\mu\nu}$ is constrained.
In the non-conformal case, Eq.(\ref{4-27}) contains on the right hand side also the quadratic correction ${\cal S}_{\mu\nu}$ of the matter fields on the brane. The effective field equations, Eq.(\ref{4-15}), are then not a closed system. One needs the Bianchi equations. In fact, ${\cal E}_{\mu\nu}$ encodes corrections from the 5D graviton effects and are for the brane observer non-local. 
In our model under consideration, we have only the $T_{\mu\nu}^{(\omega)}$ term and no source terms (apart from the 5D $\Lambda_5$). But it still source the KK modes. The dilaton $\omega$ plays the role of a "scalar field". 
But we don't need the 5D equations themselves, because the solution for $N$ is the same! It is only the $\omega^{4/3}$ which represents the 5D contribution. 
There is no exchange of energy-momentum between the bulk and brane. If one applies the model to a FLRW model\cite{roy1}, then the evolution equations are very complicated. Then inhomogeneous and anisotropic effects from the 4D matter radiation distribution on the brane are source for the 5D Weyl tensor, ${\cal E}_{\mu\nu}$ then causes non-local back-reaction on the brane. One needs an approximation scheme in order to find the missing evolution equation for ${\cal E}_{\mu\nu}$.

\subsection{ The 4D and 5D equations written in dimension $n$}
One can write the differential equations for $n=5, 4$ as
\begin{equation}
\ddot\omega=-N^4\omega''+\frac{n}{\omega(n-2)}\Bigl(N^4\omega'^2+\dot\omega^2\Bigr)\label{4-28}
\end{equation}
\begin{eqnarray}
\ddot N=\frac{3\dot N^2}{N}-N^4\Bigl(N''+\frac{3N'}{r}+\frac{N'^2}{N}\Bigr)\cr
-\frac{n-1}{(n-3)\omega}\Bigl[N^5\Bigl(\omega''+\frac{\omega'}{r}+\frac{n}{2-n}\frac{{\omega'}^2}{\omega}\Bigl)+N^4\omega' N'+\dot\omega\dot N\Bigr]\label{4-29}
\end{eqnarray}
with solution 
\begin{equation}
\omega=\Bigl(\frac{a_1}{(r+a_2)t+a_3r+a_4}\Bigr)^{\frac{1}{2}n-1}, 
N^2=\frac{1}{5r^2}\frac{10a_2^3r^2+20a_2^2r^3+15a_2r^4+4r^5+C_1}{C_2(a_3+t)^4+C_3}\qquad\label{4-30}
\end{equation}
with $a_i$ some constants.
The constraint becomes
\begin{equation}
\bar\omega''=-\frac{2n}{n-2}\frac{\Lambda l\kappa^{\frac{4}{(n-2}}\xi^{\frac{n-2}{{4(n-1)}}}\bar\omega^{\frac{n+2}{n-2}}}{N^2}-\frac{\omega' N'}{N}-\frac{\omega'}{2r}+\frac{4}{n-2}\frac{\dot{\bar\omega}^2}{\bar\omega N^4}-\frac{\dot{\bar\omega}\dot N}{N^5}\label{4-31}
\end{equation}
Note that the differential equations are constraint by an equation for $\omega''$, which contains the $\Lambda$-term.
We have now  a closed system of the 4D and 5D equations.
\subsection{ Contribution from the bulk}
If one omits the contribution from the bulk, i. e., ${\cal E}_{\mu\nu}$, one obtains the solution ($d_i$ constants)
\begin{eqnarray}
\omega=\frac{1}{(t+d_3)r+d_2t +d_2d_3},\quad N^2=\frac{1}{4r^2}\frac{6d_2^2r^2+8d_2r^3+3r^4+D_1}{D_2(t+d_3)^3+D_3}.\label{4-32}
\end{eqnarray}
The solution for $N^\varphi$ becomes
\begin{equation}
N^\varphi =\frac{(t+d_3)^2}{2r^2}\Bigl(2r^2\ln(r)-d_2(d_2+4r)\Bigr)\label{4-33}.
\end{equation}
If we compare this solution with Eq.(\ref{4-20b}), we observe that the highest power of the radial coordinate in $N^2$ is now $r^2$ in stead of $r^3$.
Further, for $D_1=0$, $N^2$ has, for any $d_2$, no real roots, so it has  naked singularities.
So the bulk contribution generates at least one horizon (see next section).
\section{Location of the horizon's, Hawking radiation and antipodal identification}
Let us inspect more closely the behavior of the induced spacetime on the brane, found in the preceding section.
It turns out, by inspection of $\bar g_{\mu\nu}$, that the spacetime  possesses horizons (coordinate singularities, i. e., 0, 1 or 2) and ergo-sphere (Killing horizon). $N^2$ becomes singular for $t=t_H=-b_3+\sqrt[4]{-\frac{C_3}{C_2}}$. However, $\bar g_{\mu\nu}$ can be made regular everywhere and singular free by suitable choices  of the parameters $b_i, c_i$ and $C_i$.
For $C_1=0$, $\bar g_{\mu\nu}$ has one real zero $r_H=\sim |1.606 b_2|$ and two complex zero's $\sim(0.178\pm 0.638I)b_2$. In figure 3 we plotted $N_1(r)^2$ for several values of $C_1$ and $b_2$. 
\begin{figure}[h]
\centerline{
\includegraphics[width=4cm]{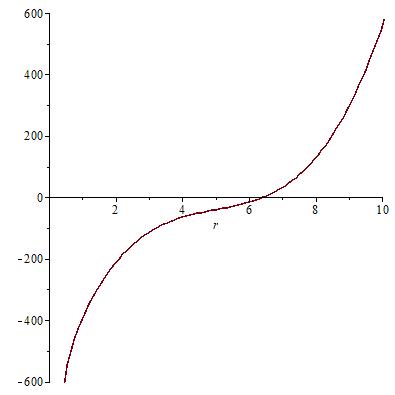}
\includegraphics[width=4cm]{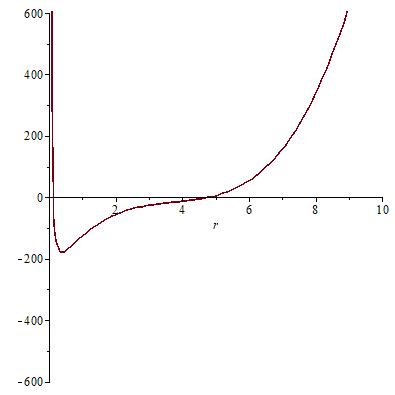}
\includegraphics[width=4cm]{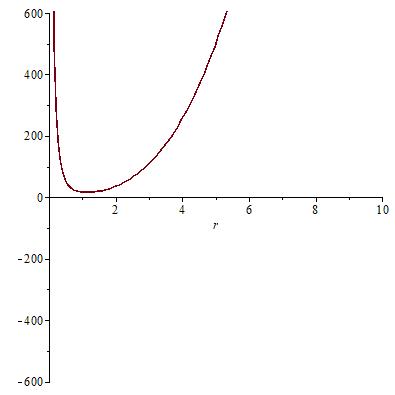}
\includegraphics[width=4cm]{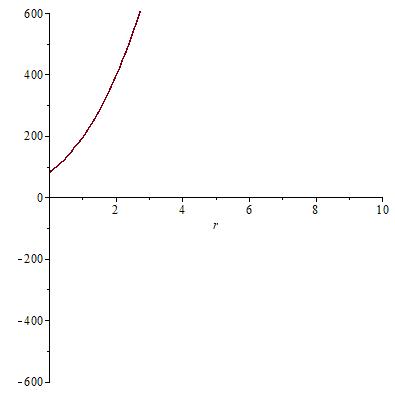}}
\caption{Four possible plots of $N_1(r)^2$.}
\end{figure}
However, there is no restriction on the angular velocity of the horizon: the equation for $N^\varphi$ decouples.
Further, one must remark that we don't deal here with the 3D BTZ (with the obligatory cosmological constant). There are no gravitational degrees of freedom in 3D Einstein gravity. Still this 3D model has interesting applications, such has the connection with the asymptotic $AdS_3$. It possesses local defects, black holes and asymptotic symmetries\cite{compere}. In our uplifted model we don't need a cosmological constant. If one omits the $dz^2$ in Eq.(\ref{4-17}), one needs a different approach\cite{slagterduston}
We shall see that the properties of our $\bar g_{\mu\nu}$ can be compared with the Kerr black hole\cite{parker}.
\subsection{The "classical" picture.}
In order to investigate the features of the time-dependent gravitational field of our rotating object  and the resulting production of Hawking radiation, one should take into account the complementarity of different observers. 
The annihilation operators of particles at late time can be expressed as superpositions of annihilation and creation operators of particles at early times (Bogolubov transformation). The created particles form a thermal spectrum determined by the surface gravity, i. e., the gravitational acceleration at the horizon. There will be also a back-reaction on the spacetime. This emission is caused by the formation of an event horizon. Finally the black hole will evaporate. During the evaporation, the surface gravity and temperature increases.

The treatment of the rotating black hole thermodynamics can be found in Wald\cite{wald}.
The mass loss can be expressed as
\begin{equation}
\delta M=\frac{\kappa}{8\pi}\delta A+\Omega_H\delta J_H.\label{5-1}
\end{equation}
with $\kappa$ the surface gravity of the black hole (see section 5.5), $A$ the area of the event horizon, $J_H$ the angular momentum and
\begin{equation}
\Omega_H=\frac{d\varphi}{dt}=-\frac{g_{t\varphi}}{g_{\varphi\varphi}}=N^\varphi\label{5-2} 
\end{equation}  
the coordinate angular velocity (the angular velocity of null  generators of the event horizon). $N^\varphi$ given by Eq.(\ref{4-11}), evaluated at $r=r_H$. It blows-up $\sim t^3$.
Massless quanta created by the black hole are governed by null geodesics that begin far outside the collapsing body at early times, move inwards and become outgoing (radially) null geodesics and escape  from the collapsing body. They reach future null infinity at arbitrarily late times. 

To describe the spectrum of the Hawking particles, one usually defines geodesics\cite{parker} $x^\mu(\lambda)$, with $\lambda$ an affine parameter. The geodesic is defined by $\nabla^\lambda(dx^\alpha/d\lambda)$, with the covariant derivative along the geodesic. In null coordinates $(u,v)$ one has for the affine separation between two outgoing null geodesics that escape from the collapsing body just before the formation of the event horizon, $u(v)=-\frac{1}{r_H}\log\frac{(v_0-v)}{K}$, with $K$ a constant.

In the Boyer-Lindquist coordinates, an incoming null geodesic take an infinite coordinate time to reach the event horizon and it will winds an infinite number of times around the z-axes.
\subsection{Changing coordinates}
If we define the coordinates, $dr^*\equiv\frac{1}{N_1(r)^2}dr$ and $dt^*\equiv N_2(t)^2dt$, then our induced spacetime of Eq.(\ref{4-17}) can be written as
\begin{equation}
ds^2=\omega^{4/3}\bar\omega^2\Bigl[\frac{N_1^2}{N_2^2}\Bigl(-dt{^*}^2+dr{^*}^2  \Bigr) +dz^2+r^2(d\varphi+\frac{N^\varphi}{N_2^2} dt^*)^2 \Bigr],\label{5-3}
\end{equation}
with
\begin{eqnarray}
N_1^2=\frac{10b_2^3r^2+20b_2^2r^3+15b_2r^4+4r^5+C_1}{5r^2},
N_2^2=\frac{1}{C_2(t+b_3)^4+C_3}\label{5-4}
\end{eqnarray}
and
\begin{eqnarray}
r^*=\frac{1}{4}\sum_{r^H_i}\frac{r^H_i \log(r-r^H_i)}{(r^H_i+b_2)^3},\qquad t^*=\frac{1}{4C_2}\sum_{t^H_i}\frac{\log(t-t^H_i)}{(t^H_i+b_3)^3}\label{5-5}.
\end{eqnarray}
The sum it taken over the roots of $(10b_2^3r^2+20b_2^2r^3+15b_2r^4+4r^5+C_1)$ and $C_2(t+b_3)^4+C_3$, i. e., $r^H_i$ and $t^H_i$. This coordinate system is comparable with a Kruskal-Szekeres coordinate system.
Further, one can define the azimuthal angular coordinate $d\varphi^*\equiv (d\varphi +\frac{N^\varphi}{N_2^2} dt^*)$, which can be used when an incoming null geodesic falls into the event horizon. $\varphi^*$  is the azimuthal angle in a coordinate  system rotating about the z-axis relative to the Boyer-Lindquist coordinates.
\subsection{Penrose diagram}
If we introduce outgoing and ingoing radial null coordinates, $u=t^*-r^*, v= t^*+r^*$, then
$u=cst.,$ and $v=cst.,$ describe the outgoing and ingoing null curves. We then have
\begin{equation}
ds^2=\omega^{4/3}\bar\omega^2\Bigl[\frac{N_1^2}{N_2^2}dudv  +dz^2+r^2 {d\varphi^*}^2 \Bigr].\label{5-6}
\end{equation}
Next we apply the usual Penrose compactification, 
\begin{eqnarray}
u=\tan U,\qquad v=\tan V.\label{5-7}
\end{eqnarray}
We then obtain
\begin{equation}
ds^2=\omega^{4/3}\bar\omega^2\Bigl[\frac{N_1^2}{N_2^2}\frac{dUdV}{\cos^2 U\cos^2 V}+dz^2+r^2{d\varphi^*}^2\Bigr].\label{5-8}
\end{equation}
Further, from Eq.(\ref{5-5}) we have
\begin{eqnarray}
r^*=\frac{1}{2}(v-u)=\frac{1}{2}(\tan V-\tan U)= \frac{1}{4}\sum_{r^H_i}\frac{r^H_i \log(r-r^H_i)}{(r^H_i+b_2)^3}    \cr
t^*=\frac{1}{2}(v+u)=\frac{1}{2}(\tan V+\tan U)=\frac{1}{4C_2}\sum_{t^H_i}\frac{\log(t-t^H_i)}{(t^H_i+b_3)^3}.\label{5-9}
\end{eqnarray}
The Penrose diagram bears  strong resemblance with  the 3D  BTZ black hole solution. However, there are  differences. First of all, we have no cosmological constant and we have have a physical 4D spacetime. Further, the factor $\frac{N_1^2}{N_2^2}$ is quite different.  $\frac{N_1^2}{N_2^2}$ can have no roots, so no horizons. Even for $r=0$, $\bar g_{\mu\nu}$ can be regular and is conformally flat ($z, \varphi$ suppressed).
This can also be seen, when one solves from Eq.(\ref{5-9}) (for the case of one horizon $r_H\sim 1.606 b_2$).
\begin{eqnarray}
r=r_H+exp\Bigl[2(v-u)(r_H+b_2)^3/rH  \Bigr],\cr
t=t_H+exp\Bigl[2C_2(v+u)(t_H+b_3)^3  \Bigr],\label{5-10}
\end{eqnarray}  
So $\bar g_{\mu\nu}$ can be used to locate the light cones.

We can write the spacetime $\bar g_{\mu\nu}$  in the form
\begin{flalign}
\bar{ds}^2=(C_2(t+b_3)^4+C_3)\Bigl[\frac{10b_2^3r^2+20b_2^2r^3+15b_2r^4+4r^5+C_1}{5r^2}dudv\Bigr]+dz^2+r^2d{\varphi^*}^2,\label{5-11}
\end{flalign}
or
\begin{eqnarray}
\bar ds^2=\Theta^2(U,V)dUdV+dz^2+r^2d{\varphi^*}^2\cr
\Theta^2=C_2\frac{e^{6(\tan V-\tan U)(r_H+b_2)^3/rH+8C_2(\tan V+\tan U)(t_H+b_3)^3}}{\cos^2 U\cos^2 V}\label{5-12}
\end{eqnarray}
Finally, we transform to coordinates $T$, $T=U+V, R=U-V$, in order to plot the Penrose diagram of figure 4.
We took $C_1=C_3=0$ and  one horizon $r_H$. We then have
\begin{eqnarray}
u=\tan U=\log \frac{(t-t_H)^\beta}{(r-r_H)^\alpha},\qquad v=\tan V=\log (t-t_H)^\beta (r-r_H)^\alpha.\label{5-7}
\end{eqnarray}
with 
\begin{equation}
\alpha=\frac{r_H}{4(r_H+b_2)^3}, \qquad \beta=\frac{1}{4C_2(t_H+b_3)^3}
\end{equation}
\begin{figure}[h]
\centerline{
\includegraphics[width=10cm]{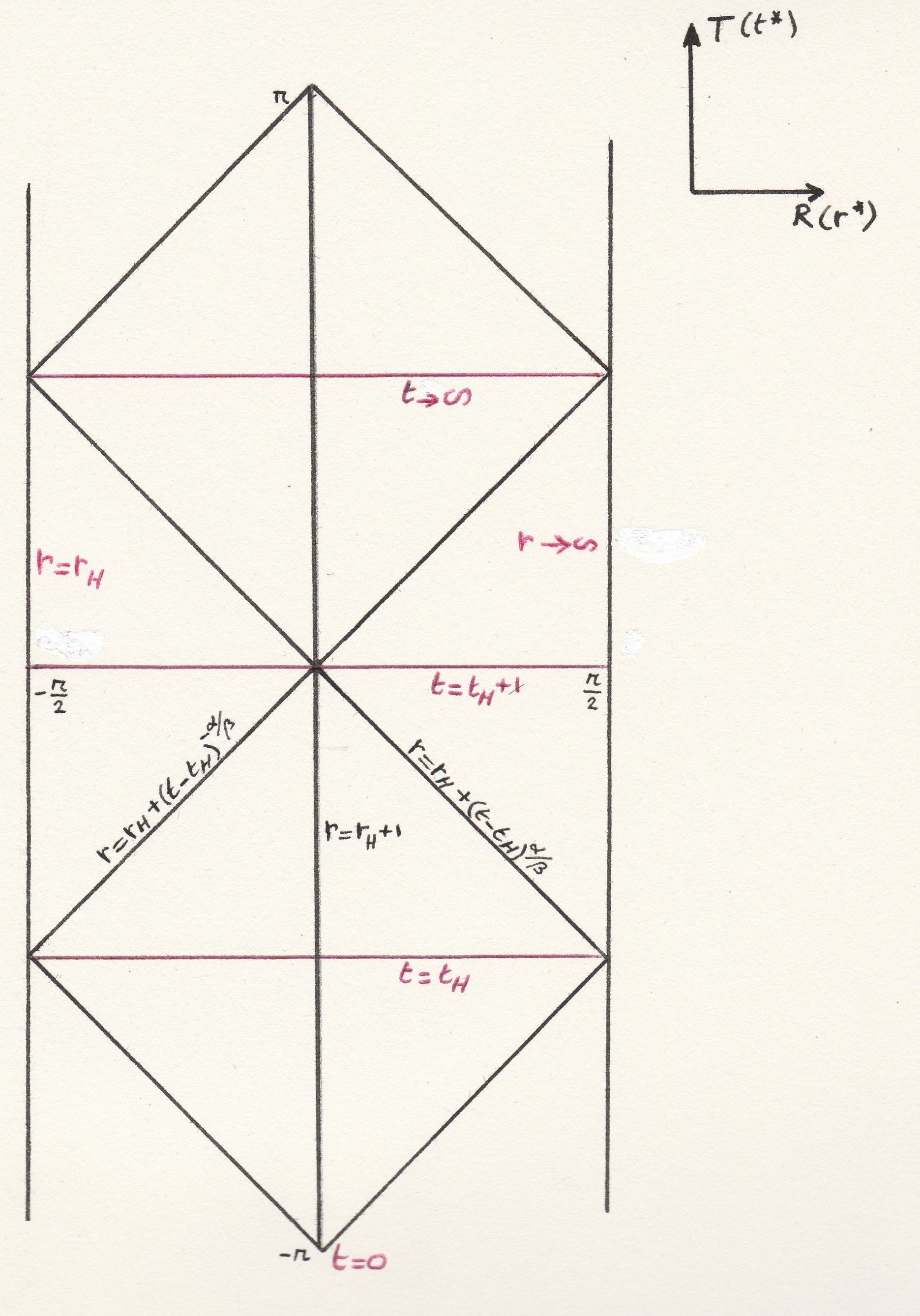}}
\caption{Kruskal diagram for $\bar g_{\mu\nu}$ in $(T,R)$-coordinates.}
\end{figure}

Note that we have the "scale-factor" $\Theta^2$ in front of $(-dT^2+dR^2)$, which causes no harm in construction the diagram. For $t=t_H$, $\Theta\rightarrow \infty$! For $t\rightarrow \infty$, $\Theta\rightarrow \infty$ OR $r\rightarrow r_H$ 
Note that although $r=0 $ has disappeared as singular point, it is in the line element Eq.(\ref{5-11}) for $C_1\neq 0$.

If one makes the antipodal identification $U\rightarrow -U, V\rightarrow -V$, one obtains $\alpha\rightarrow -\alpha$ and $\beta\rightarrow -\beta$. This is only possible if $(b_2, b_3)\rightarrow (-b_2, -b_3)$ and $r_H$ and $-r_H$ are identified! 

Following Strauss, et. al.\cite{straus}, we can use a slightly different coordinate system, in order to study quantum effects more clearly (again  1 horizon and $C_1=C_3=0$). 
\begin{figure}[h]
\centerline{
\includegraphics[width=10cm]{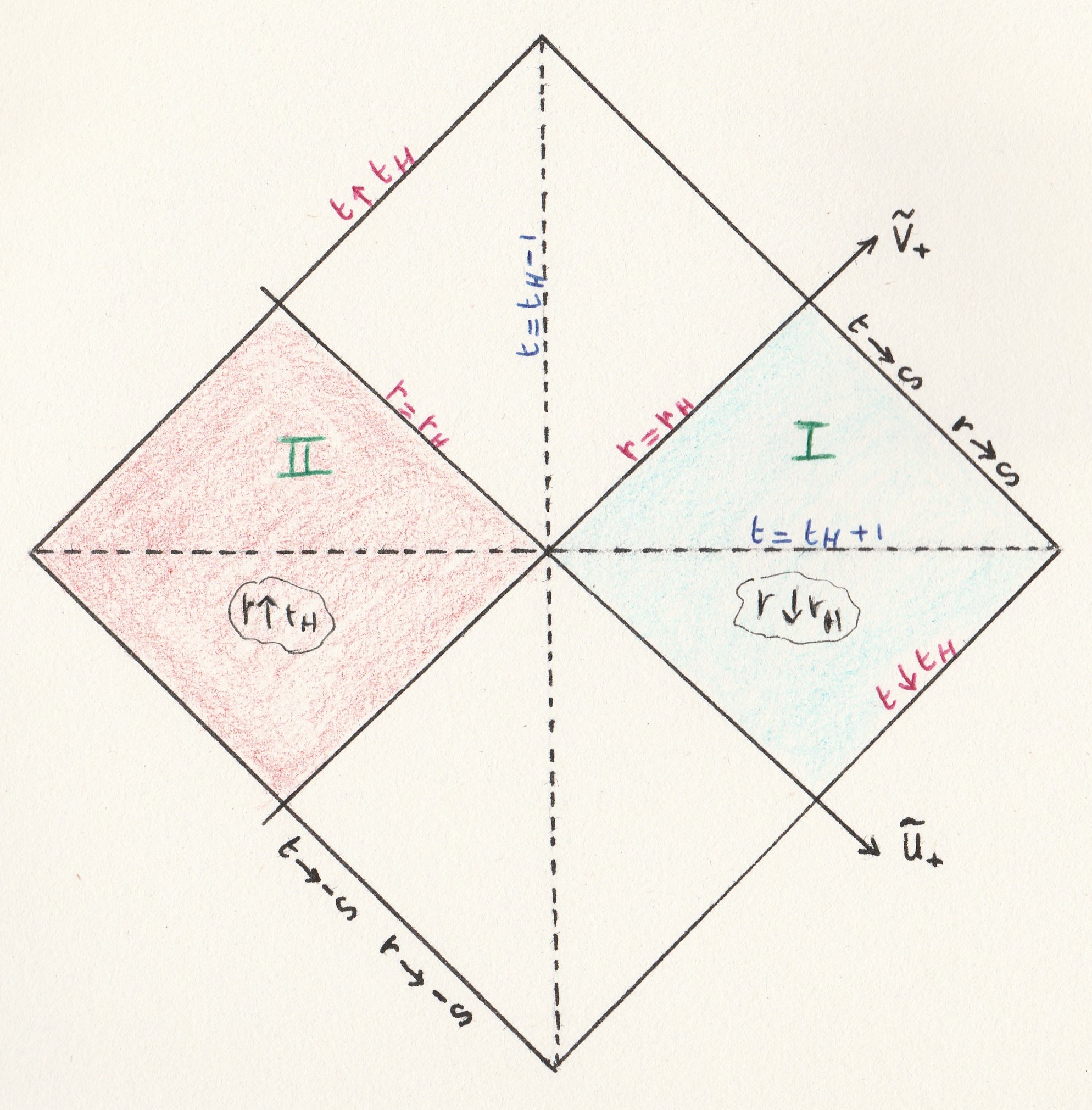}}
\caption{Kruskal diagram for $\bar g_{\mu\nu}$ in $(\tilde U, \tilde V)$- coordinates. The antipodal map between region I and II is quite clear here. If one approaches the horizon from the outside and passes the horizon, one approaches from the "otherside" the horizon. }
\end{figure}
We define 
\begin{eqnarray}
U=e^{\kappa (r^*-t^*)}, \quad V=e^{\kappa (r^*+t^*)} \qquad r>r_H \cr
U=-e^{\kappa (r^*-t^*)}, \quad V=-e^{\kappa (r^*+t^*)} \qquad r<r_H
\end{eqnarray}
Further,
\begin{equation}
\tilde U=\tanh U,\qquad \tilde V=\tanh V
\end{equation}
The spacetime can then be written as
\begin{equation}
ds^2=\omega^{4/3}\bar\omega^2\Bigl[H(\tilde U,\tilde V)d\tilde U d\tilde V +dz^2+r^2 d\varphi^{*2}\Bigr]
\end{equation}
with
\begin{equation}
H=\frac{N_1^2}{N_2^2}\frac{1}{\kappa^2\arctanh \tilde U \arctanh\tilde V(1-\tilde U^2)(1-\tilde V^2)} 
\end{equation}
We can write $r$ and $t$ as
\begin{equation}
r=r_H+\Bigl(\arctanh \tilde U \arctanh \tilde V\Bigl)^{\frac{1}{2\kappa\alpha}},\quad
t=t_H+\Bigl(\frac{\arctanh \tilde V}{ \arctanh \tilde U}\Bigl)^{\frac{1}{2\kappa\beta}}
\end{equation}
In figure 5 we plotted the Penrose diagram in $(\tilde U,\tilde V)$-coordinates.
$\tilde g_{\mu\nu}$ is regular everywhere and conformally flat. The "scale-term" H is consistent with the the features of the Penrose diagram.

\subsection{The role of the dilaton field.}
Let us now return to $g_{\mu\nu}=\omega^{4/3}\bar\omega^2\bar g_{\mu\nu}$. 
In the CDG setting, the evaporation is expressed by  the complementarity transformation of $\omega$ between the in-going and outside observer. Only the outside observer will notice the outgoing Hawking radiation. The local observer notice only the in-going radiation\cite{thooft5}.

In order to apply quantum mechanics in the  vicinity of  horizon\footnote{crossing the horizon in this antipodal mapping, also interchanges creation and annihilation operators and reverse time. So there will be an entanglement between positive energy particle with the positive energy antiparticles at the antipodes. Moreover, one can deal with pure quantum states in stead of thermodynamical mixed states. This resolves the information problem\cite{thooft6}.}, one can apply an antipodal identification, a boundary condition in order to remove the inside of the horizon. In this approach one connects the regions I and II in  the Kruskal diagram of figure 5 in order to solve the unitarity  and the entanglement problems\cite{thooft6}. This procedure shed some inside how particles transmit the information they carry, across the horizon. The horizon  changes from a 2-cylinder into a projected one.
Every point $(r,t)$ corresponds with two points $(u,v), (-u,-v)$. If one identifies $(-u,-v,z,\varphi )\equiv (u,v,-z,\varphi +\pi)$ (i.e., $t^*\rightarrow -t^*, r^*\rightarrow -r^*, z\rightarrow -z, \varphi\rightarrow \varphi +\pi$), then region I and III correspond to the same black hole, so there is no inside.

For $b_4=b_3b_2$, we have now
\begin{eqnarray}
ds^2=\omega^{4/3}\bar\omega^2\Bigl[(C_2(t+b_3)^4+C_3)\frac{(10b_2^3r^2+20b_2^2r^3+15b_2r^4+4r^5+C_1)}{5r^2}\cdot \cr \qquad\qquad\qquad
\Bigl(-dt{^*}^2+dr{^*}^2\Bigr) +dz^2+r^2(d\varphi+\frac{N^\varphi}{N_2^2} dt^*)^2\Bigr] \label{5-13}.
\end{eqnarray}
Different observers use different $\omega$ (disagreement about the vacuum state  ), i.e., the scale, by adjusting the constants.
The local observer will not notify the new boundary condition of the antipodal map. Only the outside observer will become aware of the mapping. We see that
\begin{equation}
\omega^{4/3}\bar\omega^2=\frac{1}{(r+c_2)^2(t+c_3)^2)(r+b_2)^2(t+b_3)^2}\label{5-14},
\end{equation}
approaches zero for coordinate time $t\rightarrow \infty$, so $g_{\mu\nu}$  shrinks to zero. We observe that the time derivative of Eq.(\ref{5-14}) is non-zero, so we have a non-zero $T^{\omega}$ (see Eq.(\ref{3-6})). This is the energy-momentum tensor of the Hawking particles. By suitable choices of the constants one can make it asymptotically regular.
\subsection{Surface gravity}
In order to calculate the surface gravity (needed for the Hawking radiation and the notion of the quasi-local mass)), one must realize that we are dealing here with a non-stationary black hole. For stationary Kerr black holes, one defines the Killing vector $\chi^\mu =\xi_{(t)}^\mu+\Omega_H\xi_{(\varphi)}^\mu$, normal to the Killing horizon{\footnote{ The Killing vector is a non-affinely parameterized geodesic on the horizon}. $\Omega_H$ is the angular velocity of the horizon, given by Eq.(\ref{5-2}) and $\xi_{(t)}^\mu\xi_{(t)\mu}=g_{tt}, \xi_{(\varphi)}^\mu\xi_{(\varphi)_\mu}=g_{\varphi\varphi}$. The surface gravity is then defined as 
\begin{equation}
\kappa^2=-\frac{1}{2}\nabla^\mu\chi^\nu\nabla_\mu\chi_\nu,\label{5-15}
\end{equation}
evaluated at the horizon $r=r_H$. An equivalent form is $\chi^\mu \nabla_\mu \chi^\nu=2\kappa\chi^\nu$. We say that $\chi^\mu$ is geodesic on the horizon and $\kappa$ measures the extent to which $\chi^\mu$ fails to be affinely parameterized.

In our non-stationary case, the calculations complicates, because we have an evolving horizon\cite{visser,wald}\footnote{An  approved method is the use of Eddington-Finkelstein coordinates. An other possible way out is to apply the transformation $ t\rightarrow iz, z\rightarrow it$ and follow the method of Medved, et al.\cite{med}}. 

We need now null generators of the local non-Killing horizon and a suitable normalization (parameterization) of the surface gravity. This is related to non-affinity of null geodesics on the horizon.
For a Killing vector, it is easy, because the Killing equation $\nabla_\mu l_\nu+\nabla_\nu l_\mu =0$ possesses a gauge freedom, i. e., a constant factor which can be used to re-scale $l_\mu$ and make $l^\mu l_\mu =-1$ at spacelike infinity. The Lie derivative then tells us how a spacetime tensor changes when one takes an infinitesimal step along along the vector field. For a Killing field $\chi^\mu$ we have ${\cal L}_\chi g_{\mu\nu}=0$.

Let us define  outgoing and ingoing null normals for $\bar g_{\mu\nu}$, $\bar l^\mu$ and $ \bar m^\mu$ ($\bar l^\mu \bar l_\mu = \bar m^\mu \bar m_\mu =0$, no  Killing fields), such that $\bar l^\mu \bar m_\mu=-1$:
\begin{equation}
\bar l^\mu=(1,N\sqrt{N^2-r^2{N^\varphi}^2},0,0),\quad \bar m^\mu=\Bigl(-\frac{1}{2r^2{N^\varphi}^2-2N^2},-\frac{N}{2\sqrt{N^2-r^2{N^\varphi}^2}},0,0\Bigr)\label{5-16}
\end{equation}

The surface gravity is then\cite{niels,fodor}
\begin{equation}
\kappa=-\bar m^\nu \bar l^\mu\bar\nabla_\mu \bar l_\nu.\label{5-17}
\end{equation}
One can  evaluate Eq.(\ref{5-17}) at the horizon, because we have exact expressions for $N$ and $N^\varphi$. It turns out for suitable parameters, that at $t=-b_3$, $\kappa\rightarrow \infty$.

The value of $\bar l^\mu \bar m^\nu\bar\nabla_\nu \bar m_\mu$ depends on the normalization for $l^\mu$. One should take 
\begin{equation}
\bar l^\mu \bar m^\nu\bar\nabla_\nu \bar m_\mu=0,\label{5-18}
\end{equation}
which yields an equation for $\partial_t N^\varphi$
\begin{equation}
\partial_t N^\varphi =\frac{-1}{r^2N^\varphi N}(r^2N^\varphi-2N^2)\partial_tN\label{5-19}.
\end{equation}
This determines the function of $F_2(t)$ in Eq(\ref{4-26}). This choice is also necessary, in order to obtain the correct expression for the Lie derivative ${\cal L}_{\bar l}\bar g_{\mu\nu}$

The surface gravity  then becomes
\begin{equation}
\kappa=2N\Bigl(\partial_r(\sqrt{N^2-r^2{N^\varphi}^2})+\partial_t\Bigl(\frac{1}{N}\Bigr)\Bigr)
=2N\Bigl(\partial_t\sqrt{\bar g_{rr}}-\partial_r\sqrt{\bar g_{tt}}\Bigr)\label{5-20}
\end{equation}
This is consistent with the metric definition of $\kappa$ for the  time independent case. 
\subsection{The choice of $\Omega$}
If $v^\mu$ is a vector tangent to a geodesic with affine parameter $\lambda$, then
\begin{equation}
\partial_\lambda(v^\mu l_\mu )=v^\nu\nabla_\nu(l_\mu v^\mu)=v^\nu v^\mu\nabla_\nu l_\mu=\frac{1}{4}v^\nu v_\nu\nabla^\mu l_\mu,\label{5-28}
\end{equation}
which vanishes for $v^\nu v_\nu =0$. So $l^\mu v_\mu$ is conserved along null geodesics. 
Remember that we still have  the freedom of Eq.(\ref{3-3}). One can verify that $v^\nu\nabla_\nu v^\mu$ transforms as
\begin{equation}
v^\nu\nabla_\nu v^\mu\rightarrow v^\nu\nabla_\nu v^\mu +\frac{1}{\Omega}(2v^\mu v^\nu\nabla_\nu\Omega-v^\nu v_\nu\nabla^\mu\Omega).\label{5-29}
\end{equation}
For $v^\mu v_\mu=0$ we obtain
\begin{equation}
v^\nu\nabla_\nu v^\mu\rightarrow v^\nu\nabla_\nu v^\mu +\frac{2}{\Omega} v^\mu v^\nu\nabla_\mu\Omega.\label{5-30}
\end{equation}
Now $v^\mu$ is a tangent vector to a geodesic if $v^\nu\nabla_\nu v^\mu =\alpha v^\mu$. If $\alpha =0$, we have a affinely parameterized geodesic. Conformal transformations preserve  null geodesics. So conformal transformations preserve affinely parameterized geodesics, if we take 
$\frac{2}{\Omega}v^\nu\nabla_\nu\Omega =0$, or
\begin{equation}
\partial_t\Omega =\pm N\sqrt{N^2-r^2 N^{\varphi 2}}\partial_r\Omega,\label{5-31}
\end{equation}
which is consistent with the null condition for spacetime $\tilde g_{\mu\nu}$.
It makes the surface gravity also conformal invariant (note that $\nabla_\mu \Omega^2$ is non-singular at the horizon).

We conjecture that this constraint can be used for the outside observer.
Can one construct from  $\bar l$ and $\bar m$ a conformal Killing vector? 
(note that we have ${\cal L}_{\bar l}\bar g_{\mu\nu}\neq 0$ and ${\cal L}_{\bar m}\bar g_{\mu\nu}\neq 0$)
If so, then one can construct a true Killing field for the conformally related  metric $\Omega^2\bar g_{\mu\nu}$ and a conformal invariant surface gravity  of a conformal Killing horizon. Remember, we have the conformal Killing equation for a conformal Killing vector (not null)
\begin{equation}
\bar\nabla_\mu \bar l_\nu+\bar\nabla_\nu \bar l_\mu=\frac{1}{4}\bar\nabla^\sigma \bar l_\sigma \bar g_{\mu\nu}.\label{5-32}
\end{equation}

The Lie derivative transforms as
\begin{equation}
{\cal L}_{\bar l}\bar g_{\mu\nu}\rightarrow {\cal L}_{\bar l}\log \Omega^2 \bar  g_{\mu\nu}.\label{5-33}
\end{equation}

Suppose that $\hat g_{\mu\nu}=\Omega^2\bar g_{\mu\nu}$. A spacetime with a conformal killing field $\bar l$  is conformal to a spacetime for which $\bar l$ is a true Killing field. That is, if ${\cal L}_{\bar l}\hat g_{\mu\nu}=2k\hat g_{\mu\nu}$, then ${\cal L}_{\bar l}\hat g_{\mu\nu}=0$, if 
\begin{equation}
{\cal L}_{\bar l}\Omega^2+2k\Omega^2=0.\label{5-34}
\end{equation}
The solution is given along integral curves of $\bar l$ by $\log\Omega=-\int kdv$, with $\bar l^\mu\nabla_\mu v=1$.

The black hole  complementarity is now formulated as follows: both observers uses different ways to fix the conformal gauge $\Omega$. They disagree about the vacuum state. 
In our simplified model, with exact solutions for $\omega$ and $\bar g_{\mu\nu}$, we conjecture 
that these gauges can be found by constructing  conformal  killing fields and a conformal invariant surface gravity.

\subsection{Flux density}
We can also calculate the flux density $\bar G_{tr}$. 
\begin{equation}
\frac{2(t+b_3)^3}{r\Bigl[(t+b_3)^4+\frac{C_3}{C_2}\Bigr]}.\label{5-35}
\end{equation}
Then the  flux across the horizon $r=r_H$ in an amount of proper time can be found by integrating the expression
\begin{equation}
\int_{t_{init}}^{t_{final}}\Bigl[4\pi r^2\frac{2N(t+b_3)^3}{r\Bigl[(t+b_3)^4+\frac{C_3}{C_2}\Bigr]}dt,\label{5-36}
\end{equation}
which becomes infinite at $t_H$.

\section{Numerical solutions}
We can also solve the PDE's  numerically. In figure 6 we plotted a typical solution for the 5D case.
\begin{figure}[h]
\centerline{
\includegraphics[width=5cm]{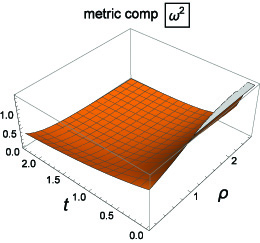}
\includegraphics[width=5cm]{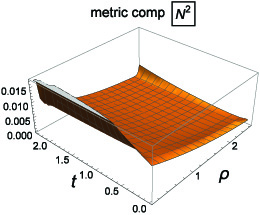}
\includegraphics[width=5cm]{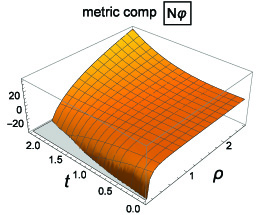}}
\centerline{
\includegraphics[width=5cm]{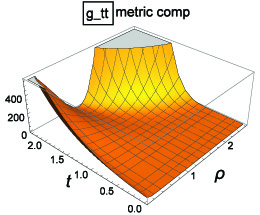}
\includegraphics[width=5cm]{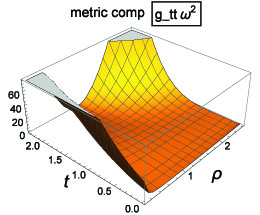}}
\caption{Example of a regular numerical vacuum solution in case of the 5D spacetime of section 4.1. We have only gravitational waves. $c_1=1; c_2=0.15;c_3=-0.2;c_4=1$.
We observe that an ergo-sphere ($g_{tt}=0$) is dynamically  formed at some $r$ and $t$. For the boundary condition for N, we used Eq(\ref{4-12}). Note that the angular momentum is depicted without an arbitrary function in t (see  Eq(\ref{4-10})). It must be chosen such that $N^\varphi$ doesn't change sign. We also see that $\omega$ shrinks gradually to zero as the $t\rightarrow\infty$ (see also the exact solution).}
\end{figure}
The numerical solution of the 4D counterpart case doesn't differ significantly. 
With the numerical "back-up" solution, one can investigate how the constants of the exact solution are related to the boundary conditions and the $\Lambda\omega^4$ term.

\section{Conclusions}
We investigated the conformal invariance of the Einstein-Hilbert action  on a dynamical warped 5D black hole spacetime. 
By solving both Einstein and dilaton equations, we find an exact solution for the dilaton field and the metric components of the "un-physical" spacetime $\bar g_{\mu\nu}$, where the real spacetime is written as $^{(5)}{g_{\mu\nu}}=\omega^{4/3}{^{(5)}{\tilde g_{\mu\nu}}} $ and  $ ^{(4)}{\tilde g_{\mu\nu}}=\bar\omega^2{ ^{(4)}\bar g_{\mu\nu}}$.
In our model, $\omega$ can be seen as the contribution of gravitational waves from the bulk (KK-modes), while $\bar \omega$ is the brane component. There is no exchange of energy-momentum between the bulk and brane in our vacuum model. 

It is conjectures that the different conformal gauge freedom, $\Omega$, the in-going and outside observers possesses, can be calculated by demanding a conformal invariant surface gravity. This means that the complementarity is expressed by the different notion of the vacuum state. They disagree about the interior of the black hole. 
The solution guarantees  regularity of the action when $\omega \rightarrow 0$.
We don't need a Weyl term in the action (generates negative metric states). In stead, we have a contribution from the bulk, i. e., the electric part of the 5D Weyl tensor. 
It is remarkable that the 5D field equations and the effective 4D equations can be written for general dimension $n$, with $n=4,5$. 
The energy-momentum tensor of the time-dependent dilaton, determining  the Hawking radiation, can be calculated exactly.
By suitable choice of the parameters, the spacetime $\bar g_{\mu\nu}$ can be regular and singular free.

In context of quantization procedures, counter terms in an effective action will cause problems, only in the bulk spacetime of the "large" extra dimension and not for the brane spacetime. When the extra dimensional volume is significantly above the Planck scale, then the true fundamental scale can be much less than the effective scale $10^{19}$ GeV. This means that no UV cut-off is necessary on the brane.
This exact solution, nonetheless without mass terms, can be used to tackle the deep-seated problem of the black hole complementarity: the infalling and outside observer experience different $\omega$ by the choice of $\Omega$. 

The solution fits also very well in the antipodal mapping, when crossing the horizon.
The Penrose diagram for $\bar g_{\mu\nu}$, in suitable Kruskal coordinates, shows the features of  the antipodal map of region I on region II: the inside of the black hole is removed. The in-going observer, when crossing the horizon, turns up at "the other side" of the horizon.

The next task is to incorporate  mass into our model and investigate the dilaton-scalar field interaction.
The conformal invariance will then  spontaneously be broken. It is conjectured that the parameters of the model will be calculable in the final version.
This is currently under investigation by the author.

\section{References}
\end{document}